\documentclass[twocolumn,aps,prd,showpacs,groupedaddress,nofootinbib]{revtex4-1}
\usepackage{amsmath}
\usepackage{bm}
\usepackage{slashed}
\usepackage{epsfig}
\usepackage{amsfonts}
\usepackage{xcolor}
\usepackage{epstopdf}
\usepackage{enumitem}
\usepackage{subfig}
\graphicspath{{figures/}{fig/}}

\def\be{\begin{equation}}
\def\ee{\end{equation}}
\def\bea{\begin{eqnarray}}
\def\eea{\end{eqnarray}}
\def\NO{\nonumber}
\def\gev{\mathrm{~GeV}}

\begin{document}


\title{Prompt $J/\psi$ photoproduction within the non-relativistic QCD framework at the CEPC}
\author{Xi-Jie Zhan}
\email{zhanxj@ihep.ac.cn}
\author{Jian-Xiong Wang}
\email{jxwang@ihep.ac.cn}
\affiliation{
	Institute of High Energy Physics, Chinese Academy of Sciences, P.O.Box 918(4), Beijing 100049, China \\
	School of Physical Sciences, University of Chinese Academy of Sciences, Beijing 100190, China
}

\date{\today}

\begin{abstract}
The prompt $J/\psi$ photoproduction within the non-relativistic QCD (NRQCD) framework at the future Circular Electron Positron Collider (CEPC) is studied, including the contributions from both direct and resolved photons. Employing different sets of long distance matrix elements, the total cross section is dominated by the color-octet channel. We present different kinematic distributions of $J/\psi$ production and the results show there will be about 50 $J/\psi$ events when the transverse momentum of $J/\psi$ is up to 20 GeV. It renders that the $J/\psi$ photoprodution at the CEPC is a well laboratory to test the NRQCD and further clarify the universality problem in NRQCD between electron positron collider and hadron collider.
\end{abstract}

\maketitle
\allowdisplaybreaks

\section{Introduction\label{sec:introduction}}

The production and decay of heavy quarkonium are ideal processes to study both perturbative and nonperturbative aspects of quantum chromodynamics (QCD). Due to the heavy quark mass , non-relativistic QCD (NRQCD)~\cite{Bodwin:1994jh} was proposed to calculate its production and decay.
In the NRQCD factorization formalism, the production of heavy quarkonium is factorized into a product of the short distance coefficients (SDCs) and long distance matrix elements (LDMEs), where the former are process-dependent and can be perturbatively calculated as double expansions in both the coupling constant $\alpha_s$ and the heavy quark relative velocity $v$, and the latter are supposedly universal and can be determined by fitting to experimental data. 

NRQCD predicts the color-octet processes and have obtained success in $J/\psi$ production~\cite{Campbell:2007ws,Gong:2008ft,Butenschoen:2010rq,Ma:2010yw} and 
polarization~\cite{Gong:2010bk,Butenschoen:2012px,Chao:2012iv,Gong:2012ug,Feng:2018ukp} at hadron colliders.
At $e^+e^-$ colliders, heavy quarkonium have also been studied. There are advantages on both experimental and theoretical sides~\cite{He:2019tig}. On the experimental side, there is less and cleaner background for signal reconstruction; On the theoretical side, the mechanism of production is simpler and the uncertainties in calculation are smaller. There are two modes of heavy quarkonium production, i.e., $e^+e^-$ annihilation and $\gamma\gamma$ collision. The $J/\psi$ production through these two modes had been measured at B factories\cite{Abe:2001za,Aubert:2001pd} and CERN LEP-II\cite{TodorovaNova:2001pt, Abdallah:2003du} respectively. As for the results of $e^+e^-$ annihilation, NRQCD can not give congruous explanation. For example, the Belle result of $J/\psi$+$X_{non-c\overline{c}}$ production favors the CS prediction and is overshoot by the NRQCD predictions\cite{Ma:2008gq,He:2009uf}. In the $\gamma\gamma$ collision case, the leading order(LO) NRQCD calculation~\cite{Klasen:2001cu} can explain the LEP-II data but the NLO~\cite{Butenschoen:2011yh} fails. What is worthy to note is that the uncertainties of LEP-II measurements are very large. With an integrated luminosity of 617 $\mathrm{pb}^{-1}$, there are only 36$\pm$7 $J/\psi\rightarrow\mu^+\mu^-$ events and only 16 thereof are in the region $p_t > 1\gev$.

The proposed Circular Electron Positron Collider (CEPC) is a powerful $e^+e^-$ collider\cite{CEPCStudyGroup:2018rmc,CEPCStudyGroup:2018ghi}. It could be operated at center of mass energy of the Z pole (91.2 GeV), the WW threshold (161 GeV), the Higgs factory (240 GeV) and the peak luminosity at 240$\gev$ is order of $10^{34}\mathrm{cm}^{-2}\mathrm{s}^{-1}$. Consequently, enormous $J/\psi$ events and high precision measurement are expected. Under the energy of 240$\gev$, the $\gamma\gamma$ collision mode is dominant. The measurement of $J/\psi$ photoproduction at the CEPC would supply precision results of different kinematics distributions and it could hopefully clarify the current predicament. Surely, it could also deepen our understanding of the physics of heavy quarkonium.

In this work, we study the prompt $J/\psi$ photoproduction at the $e^+e^-$ collider CEPC, including both direct $J/\psi$ production and feed-down contribution from heavier quarkonium.
In Section~\ref{sec:framework}, we gives the basic NRQCD framework for the calculation. Section~\ref{sec:numerical} shows the numerical results. A brief summary and conclusion is presented in Section~\ref{sec:summary}.

\section{Prompt $J/\psi$ Photoproduction in NRQCD Framework\label{sec:framework}}

Electron-positron bremsstrahlung is the source of photons, which can be well formulated in Weiz\"acker-Williams approximation(WWA)~\cite{Frixione:1993yw},
\begin{eqnarray}
\label{eq:wwa}
f_{\gamma/e}(x) &=& \frac{\alpha_{em}}{2\pi}\Bigg[\frac{1 + (1 - x)^2}{x} {\rm log}\frac{Q^2_{max}}{Q^2_{min}} \NO\\
&&+2m_e^2x\left(\frac{1}{Q^2_{max}}
-\frac{1}{Q^2_{min}}\right)\Bigg],
\end{eqnarray}
where $\alpha_{em}=1/137$, $Q^2_{min} = m_e^2 x^2/(1-x)$ and $Q^2_{max} = (E\theta_c)^2(1-x) + Q^2_{min}$ with $x = E_\gamma/E_e$, $\theta_c=32\mathrm{~mrad}$ the maximum scattered angular cut in order to ensure the photon to be real, and $E=E_e=\sqrt{s}/2$, which is $\sqrt{s}=240\gev$ at the CEPC.

In NRQCD, the SDCs stand for the production of a pair of quark-antiquark in intermediate Fock state $n={}^{2S+1}\!L^{[c]}_{J}$, where $S$ is total spin, $L$ is orbital angular momentum, $J$ is total angular momentum and $c=1,8$ for color-singlet (CS) and color-octet (CO) state respectively. The LDMEs describe the probability of the intermediate state evolving into the meson.
The colliding photons can participate in the hard interaction either directly or via their hadronic components. Under the picture of WWA and the factorization of NRQCD, the overall differential cross section of a Hadron($H$) photoproduction can be obtained by the double convolution of the cross sections of parton-parton (photon-photon) and parton distribution functions.
\begin{eqnarray}
\label{tcs}
&&d\sigma(e^+e^-\to e^+e^-H+X)\NO\\
&&~=\int dx_1f_{\gamma/e}(x_1)\int dx_2f_{\gamma/e}(x_2)\NO\\
&&~~~~\times\sum_{i,j,k}\int dx_if_{i/\gamma}(x_i,\mu_f)\int dx_jf_{j/\gamma}(x_j,\mu_f)\nonumber\\
&&~~~~\times\sum_nd\sigma(ij\to c\overline{c}[n]+k)\langle{\cal O}^H[n]\rangle,
\end{eqnarray}
here $f_{i/\gamma}(x)$ denotes the the Gl\"uck-Reya-Schienbein parton distribution functions in photon~\cite{Gluck:1999ub},
$d\sigma(ij\to c\overline{c}[n]+k)$ are the differential partonic cross sections, $i,j=\gamma,g,q,\bar{q}$ and $k=g,q,\bar{q}$ with $q=u,d,s$. $\langle{\cal O}^H[n]\rangle$ are the LDMEs of $H$.
$c\overline{c}[n]$ are the intermediate $c\overline{c}$ Fock states with $n={}^3\!S_1^{(1)},{}^1\!S_0^{(8)},{}^3\!S_1^{(8)},{}^3\!P_J^{(8)}$ for $H=J/\psi,\psi(2S)$ and $n={}^3\!P_J^{(1)},{}^3\!S_1^{(8)}$ for $H=\chi_{cJ}$ where $J=0,1,2$.

$J/\psi$ mesons can be produced directly or via decays of heavier charmonia, such as $\psi(2S)$ and $\chi_{cJ}(J=0,1,2)$. These feed-down contributions to $J/\psi$ production can be included by multiplying their direct-production cross sections with their decay branching ratios to $J/\psi$, i.e.,
\begin{eqnarray}
\label{feeddown}
d\sigma^{\mathrm{prompt}J/\psi}&=&d\sigma^{J/\psi}+\sum_{J}d\sigma^{\chi_{cJ}}Br(\chi_{cJ}\rightarrow J/\psi+\gamma)\nonumber\\
&&+d\sigma^{\psi(2S)}Br(\psi(2S)\rightarrow J/\psi+X).
\end{eqnarray}

\begin{figure}[!h]
	\includegraphics[scale=0.8]{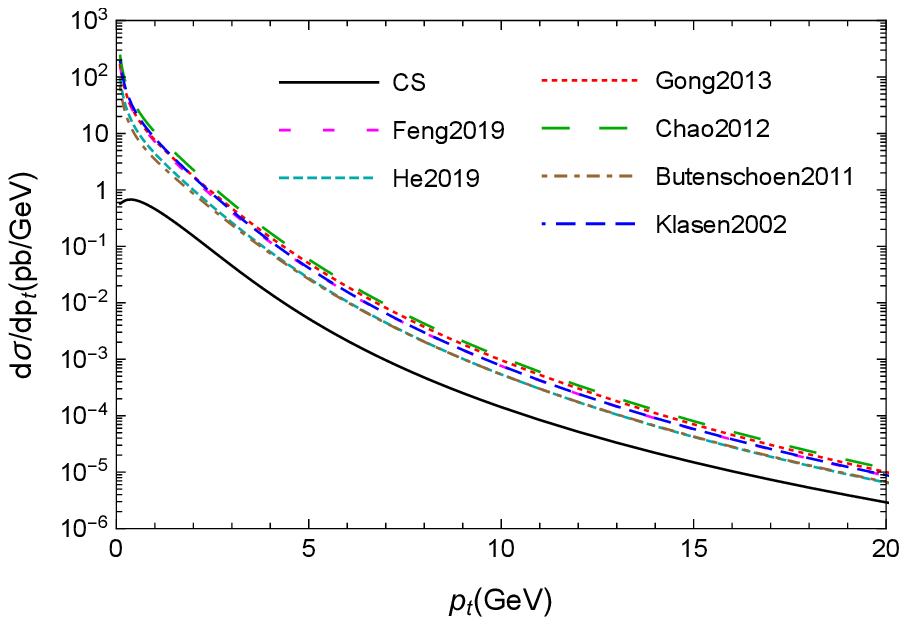}\\
	\includegraphics[scale=0.8]{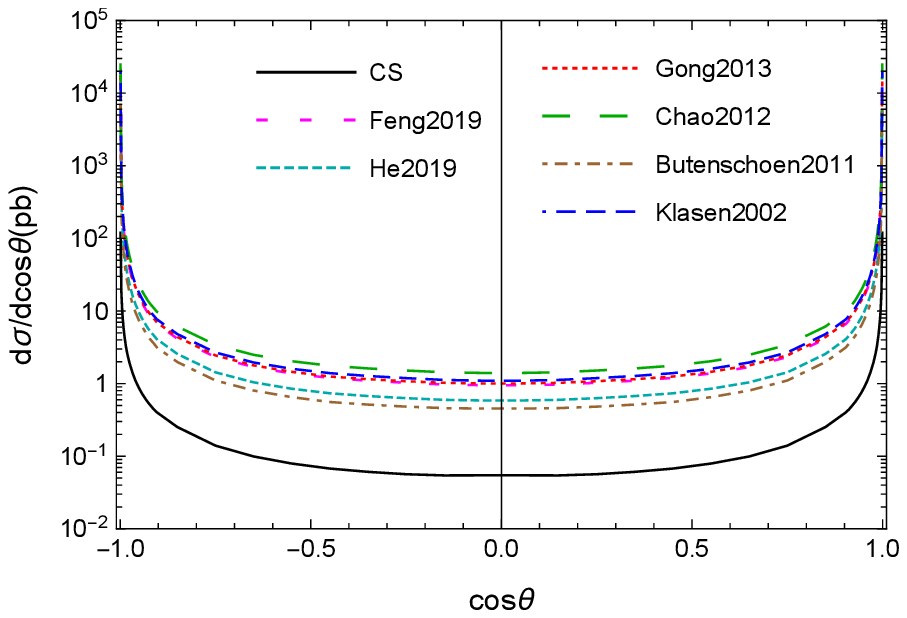}\\
	\includegraphics[scale=0.8]{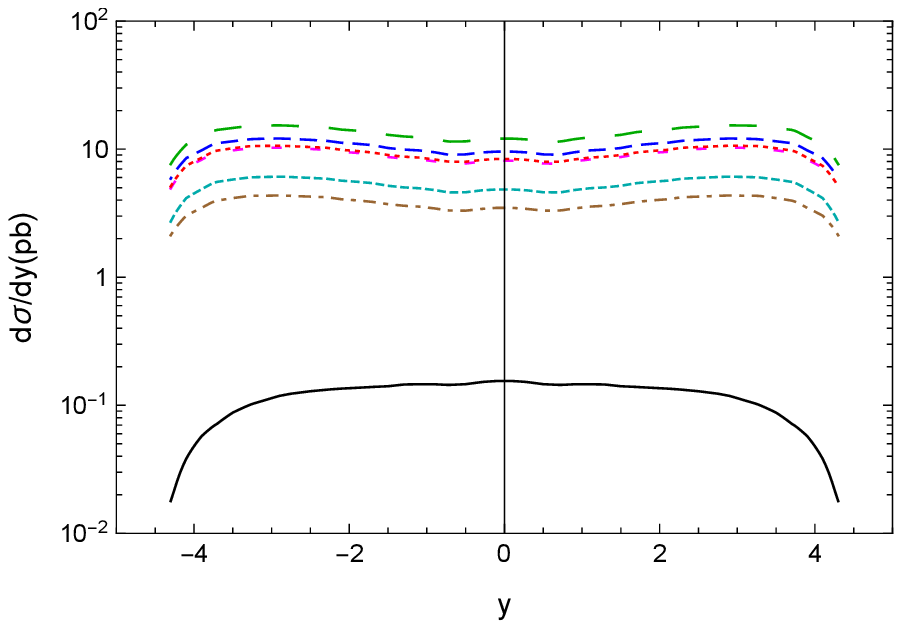}
	\caption{\label{fig:distr1}
		The $p_t$(upper), cos$\theta$(mid), and $y$(lower) distribution of prompt $J/\psi$ employing different CO LDMEs in Table~\ref{tab:coldmes}. The $y$ plots use same legends as $p_t$.}
\end{figure}

\begin{table*}[htb] 
	\begin{tabular*}{\textwidth}{@{\extracolsep{\fill}}l*{8}{c}}
		\hline
		\hline
		References
		& He \textit{et al}.
		& Feng \textit{et al}.
		& Gong \textit{et al}.
		& Chao \textit{et al}.
		& Butenschoen \textit{et al}.
		& Klasen \textit{et al}. 
		&\\
		& (2019)~\cite{He:2019qqr}
		& (2019)~\cite{Feng:2018ukp} 
		& (2013)~\cite{Gong:2012ug}
		& (2012)~\cite{Chao:2012iv} 
		& (2011)~\cite{Butenschoen:2011yh}
		& (2002)~\cite{Klasen:2001cu} 
		&\\
		\hline
		$\langle\mathcal{O}^{J/\psi}({}^1\!S_0^{[8]})\rangle$ &
		$3.61\times 10^{-2}$&
		$5.66\times 10^{-2}$&
		$9.7\times 10^{-2}$&
		$8.9\times 10^{-2}$&
		$3.04\times 10^{-2}$&
		$4.35\times 10^{-2}$&
		\\
		$\langle\mathcal{O}^{J/\psi}({}^3\!S_1^{[8]})\rangle$ &
		$1.25\times 10^{-3}$&
		$1.77\times 10^{-3}$&
		$-4.6\times 10^{-3}$&
		$3.0\times 10^{-3}$&
		$1.68\times 10^{-3}$&
		$4.4\times 10^{-3}$&
		\\
		$\langle\mathcal{O}^{J/\psi}({}^3\!P_0^{[8]})\rangle/m_Q^2$ &
		$0$&
		$3.42\times 10^{-3}$&
		$-9.5\times 10^{-3}$&
		$5.6\times 10^{-3}$&
		$-4.04\times 10^{-3}$&
		$1.28\times 10^{-2}$&
		\\
		$\langle\mathcal{O}^{\psi(2S)}({}^1\!S_0^{[8]})\rangle$ &
		$2.19\times 10^{-2}$&
		$-1.2\times 10^{-4}$&
		$-1.2\times 10^{-4}$&
		$0$&
		$0$&
		$6.5\times 10^{-3}$&
		\\
		$\langle\mathcal{O}^{\psi(2S)}({}^3\!S_1^{[8]})\rangle$ &
		$3.41\times 10^{-4}$&
		$3.4\times 10^{-3}$&
		$3.4\times 10^{-3}$&
		$0$&
		$0$&
		$4.2\times 10^{-3}$&
		\\
		$\langle\mathcal{O}^{\psi(2S)}({}^3\!P_0^{[8]})\rangle/m_Q^2$ &
		$0$&
		$4.2\times 10^{-3}$&
		$4.2\times 10^{-3}$&
		$0$&
		$0$&
		$1.86\times 10^{-3}$&
		\\
		$\langle\mathcal{O}^{\chi_{c0}}({}^3\!S_1^{[8]})\rangle$ &
		$5.29\times 10^{-4}$&
		$2.21\times 10^{-3}$&
		$2.21\times 10^{-3}$&
		$0$&
		$0$&
		$2.3\times 10^{-2}$&
		\\
		\hline
		\hline
	\end{tabular*} 
	\caption{Different sets of CO LDMEs (in units of GeV$^3$).}
	\label{tab:coldmes} 
\end{table*}

\section{Numerical Results\label{sec:numerical}}

For all the physics processes, the Fortran source for numerical calculation are generated by using the FDC package\cite{Wang:2004du}. In the numerical calculations, we take $\alpha=1/128$ and one-loop running $\alpha_s(\mu_r)$.
The charm-quark mass is approximately chosen as $m_c=m_H/2$, where the relevant quarkonia masses are $m_H=3.097,3.415,3.511,3.556,3.686\gev$ for $H=J/\psi,\chi_{cJ}(J=0,1,2)$ and $\psi(2S)$, respectively.
The renormalization scale($\mu_r$) and factorization scale($\mu_f$) are set to be $\mu_r=\mu_f=\sqrt{4m_c^2+p_t^2}$, where $p_t$ is the transverse momentum of $H$ meson.
The relevant branching ratios are $Br(\psi(2S)\rightarrow J/\psi)$=0.61 and
$Br(\chi_{cJ}\rightarrow J/\psi)$=0.014, 0.343, 0.19
for $J=0,1,2$~\cite{Tanabashi:2018oca}, respectively. The integrated luminosity of CEPC is $5.6~\mathrm{ab}^{-1}$~\cite{CEPCStudyGroup:2018ghi}.
Additionally, a shift $p_t^H\approx p_t^{H'}\times(M_H/M_{H'})$ is used while considering the
kinematics effect from higher excited states. As a check, we have reproduced the results relevant with LEP-II in Ref.\cite{Klasen:2001cu,Butenschoen:2011yh}.

\begin{figure}
	\includegraphics[scale=0.8]{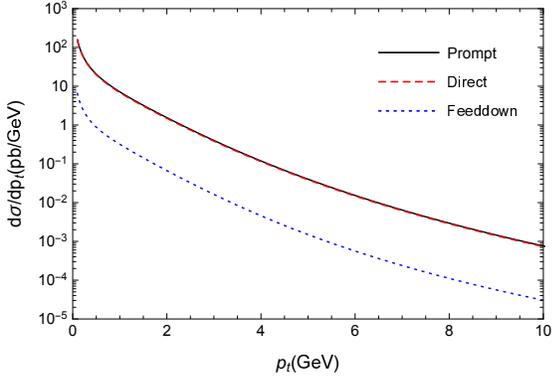}\\
	\caption{\label{fig:feeddown}
		The $p_t$ distributions of direct $J/\psi$ production and that from heavier charmonia.}
\end{figure}
\begin{figure}
	\includegraphics[scale=0.8]{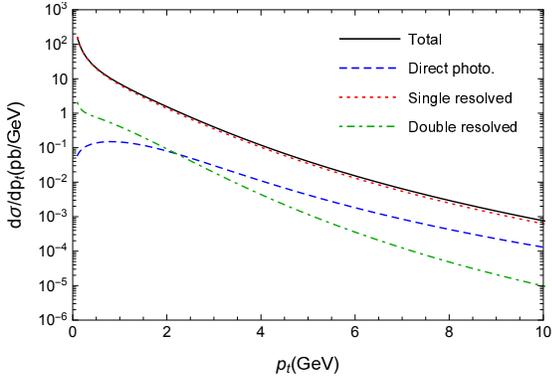}\\
	\caption{\label{fig:resolved}
		The $p_t$ distributions of the cross section from direct photoproduction and resolved photoproduction.}
\end{figure}
\begin{figure}
	\includegraphics[scale=0.8]{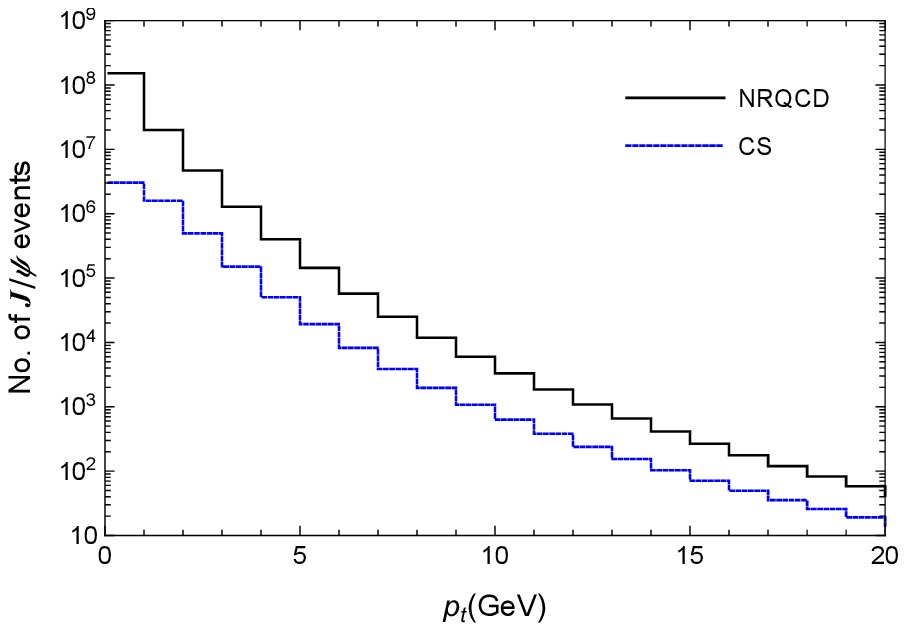}\\
	\includegraphics[scale=0.8]{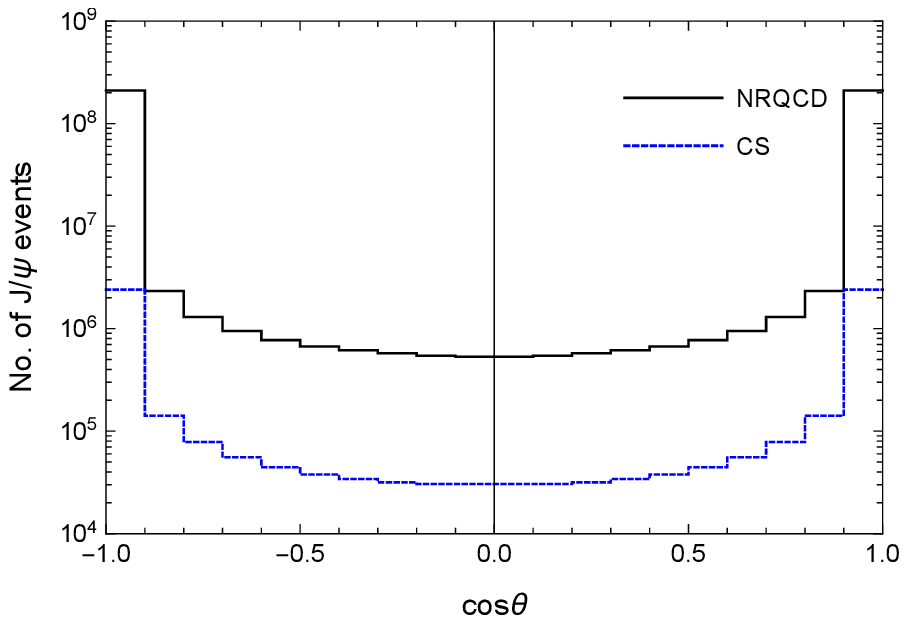}
	\caption{\label{fig:event}
		The $p_t$ and $\cos\theta$  distributions of the number of $J/\psi$ events. The bin widths are 1$\gev$ for $p_t$ and 0.1 for $\cos\theta$.}
\end{figure}

\begin{figure}
	\includegraphics[scale=0.8]{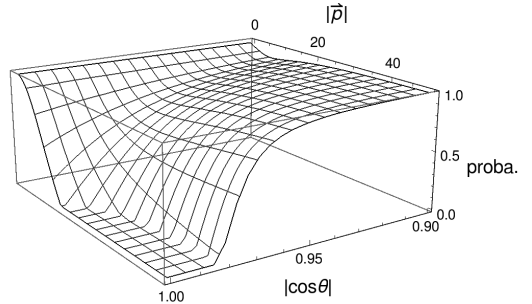}\\
	\caption{\label{fig:proba}
		The probability distribution of $J/\psi$ meson with momentum $\vec{p}$.}
\end{figure}

\begin{figure}
	\includegraphics[scale=0.7]{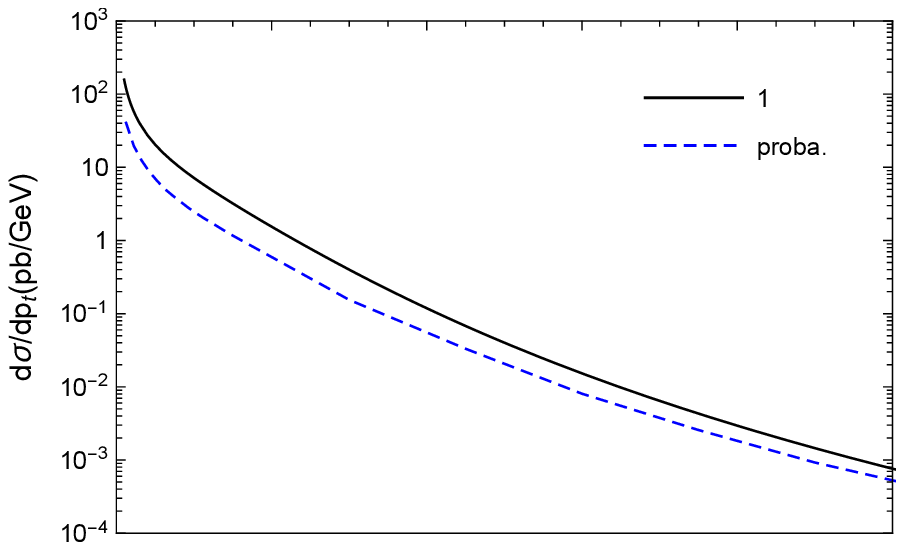}\\ \hspace{0.98mm}	
	\includegraphics[scale=0.7]{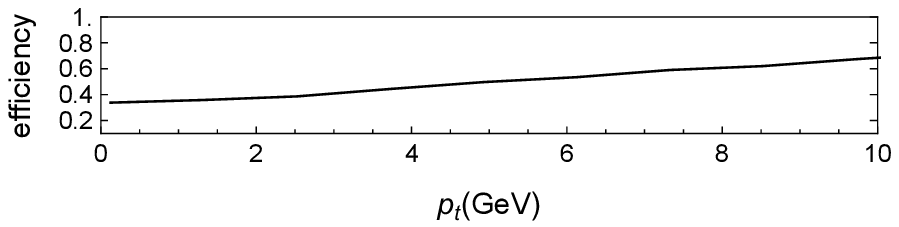}\\ \vspace{2mm}
	\includegraphics[scale=0.7]{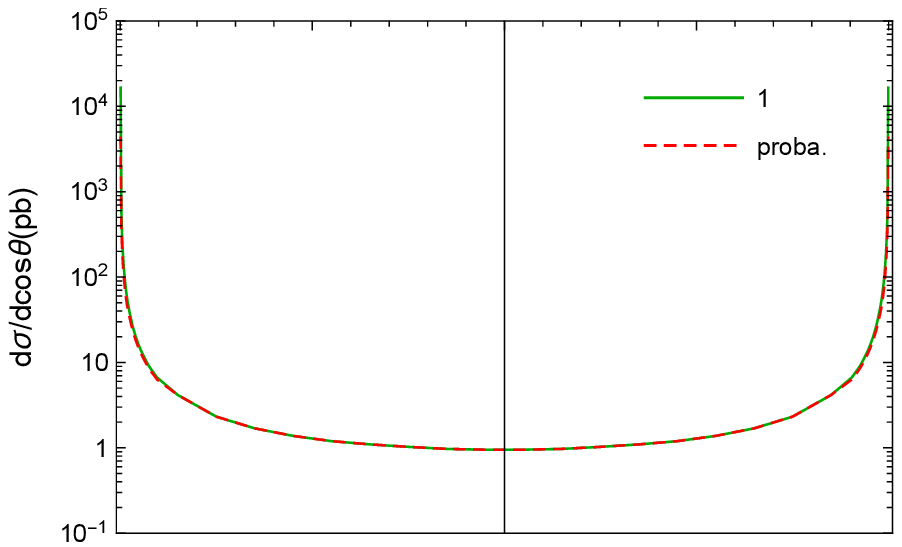}\\ \hspace{0.98mm}	
	\includegraphics[scale=0.7]{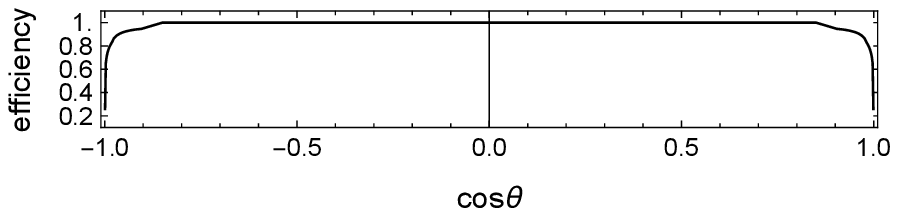}\\ \vspace{2mm}
	\includegraphics[scale=0.7]{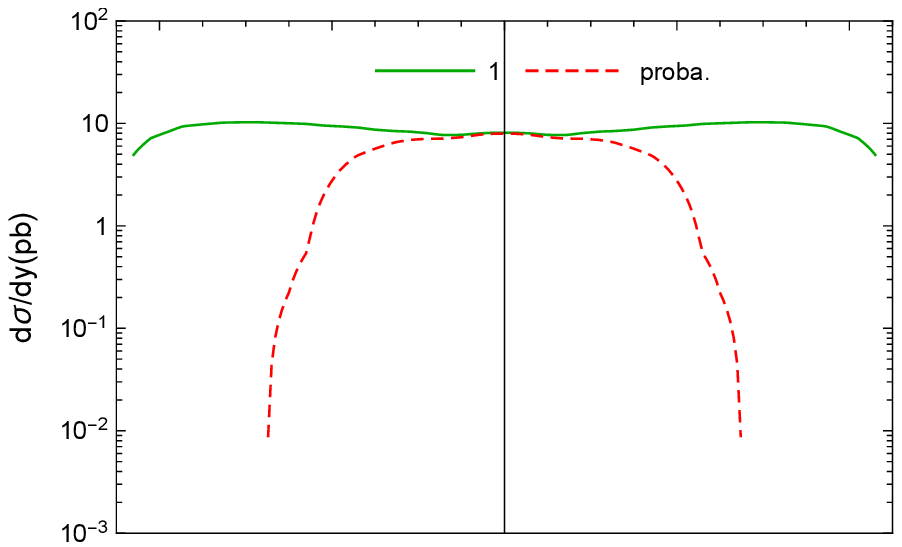}\\ \hspace{-1.1mm}	
	\includegraphics[scale=0.7]{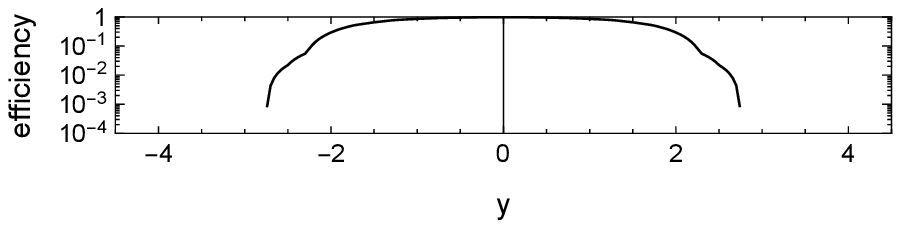}
	\caption{\label{fig:kinproba}
		The kinematic distributions of prompt $J/\psi$ photoproduction before(Line-1) and after(Line-proba.) considering the detection efficiency. The curves in the flat frames are the corresponding efficiencies.}
\end{figure}

The CS LDMEs are estimated from the wave functions at the origin by
\begin{eqnarray}
\langle{\cal O}^{\psi}(^{3}S^{[1]}_{1})\rangle&=&\frac{3N_c}{2\pi}|R_{\psi}(0)|^{2}, \NO\\
\langle{\cal O}^{\chi_{cJ}}(^{3}P^{[1]}_{J})\rangle&=&\frac{3}{4\pi}(2J+1)|R'_{\chi_{c}}(0)|^{2},
\end{eqnarray}
where the wave functions at the origin can be calculated via potential model~\cite{Eichten:1995ch},
which gives $|R_{J/\psi}(0)|^{2}$=0.81$\gev^3$, $|R_{\psi(2S)}(0)|^{2}$=0.53$\gev^3$, and $|R'_{\chi_c}(0)|^{2}$=0.075$\gev^5$.

There are several sets of CO LDMEs on the market and it is instructive to compare their predictions for the $J/\psi$ photoproduction at $e^+e^-$ collision. We employ six different sets of CO LDMEs listed in Table~\ref{tab:coldmes}. Since all the LDMES are extracted at $m_c=1.5\gev$, the scaling rules need to employ,
\begin{eqnarray}
\langle\mathcal{O}^{H}({}^{2s+1}\!S_j^{[n]})\rangle&\propto&m_c^3, \NO\\
\langle\mathcal{O}^{H}({}^{2s+1}\!P_j^{[n]})\rangle/m_c^2&\propto&m_c^3.
\end{eqnarray}

Fig.~\ref{fig:distr1} presents the $p_t$, cos$\theta$ and rapidity($y$) distribution of prompt $J/\psi$ photoproduction for different CO LDMEs, where $\theta$ is the angle between $J/\psi$ and beam. Both cos$\theta$ and $y$ distribution are calculated under the cut condition $p_t\geq0.01\gev$. Unlike in the $J/\psi$ hadroproduction, it shows that these CO LDMEs give different predictions for $J/\psi$ photoproduction at the CEPC, which provides a good laboratory to scrutinize the validity of them. From the curves, the CO contributions are always larger than that of CS and distinguishable from CS results and hence make it easy to test the NRQCD color-octet mechanism.

Taking the CO LDMEs of Feng2019\cite{Feng:2018ukp} as default choice, we show the feed-down and resolved contributions of NRQCD in Fig.~\ref{fig:feeddown} and Fig.~\ref{fig:resolved}, respectively. It is clear that most of $J/\psi$ mesons are produced directly and only 5$\%$ of them are from decays of $\psi(2S)$ and $\chi_{cJ}$. The single resolved channel is also dominated. As the reference, the $p_t$ distribution integrated from 0.1 to 10$\gev$, the direct, single-resolved and double-resolved channels account for 1$\%$, 96$\%$ and 3$\%$ of the NRQCD prediction, respectively. Fig.~\ref{fig:event} shows the number of $J/\psi$ events distribution as function of $p_t$(upper) and $\cos\theta$(lower) respectively for the integrated luminosity of CEPC $5.6~\mathrm{ab}^{-1}$. We see that at the CEPC, the number of events are considerable to discriminate between CS and NRQCD even as the transverse momentum $p_t$ is up to $20\gev$.

The $\cos\theta$ distributions in Fig.~\ref{fig:distr1} shows that most of $J/\psi$ are located in closed beam region. Specifically, more than 90$\%$ of $J/\psi$ mesons are inside $|\cos\theta|\geq0.98$, the angular cut for experimental detection. In fact, $J/\psi$ mesons decay immediately after production at the colliding point. In experimental measurement $\mu^+\mu^-$, for example, is used to reconstruct $J/\psi$ meson and hence the probability of the $\mu^+\mu^-$ pair to be detected should be investigated. If both $\mu^+$ and $\mu^-$ are detected at laboratory frame, then their parent $J/\psi$ meson is a valid event. So there is an issue of detection efficiency for $J/\psi$. For simplicity we assume that, at the center-of-mass frame of $J/\psi$, the $\mu^+\mu^-$ pair are isotropic in the whole $4\pi$ solid angle. Then we can easily calculate the probability of a $J/\psi$ meson with given 4-momentum to be a valid event. Some brief derivations are given in the Appendix. In Fig.~\ref{fig:proba} we plot the two-dimension distribution of the probability as function of magnitude of 3-momentum and $|\cos\theta|$ of $J/\psi$. It shows that $J/\psi$ mesons with $|\cos\theta|\geq0.98$ but small $|\vec{p}|$ still have the probability to be a valid event. Fig.~\ref{fig:kinproba} presents three kinematic distributions both before(Line-1) and after(Line-proba.) considering the detection efficiency and here only the NRQCD results are shown. The plots show that the efficiency gets larger as $p_t$ increases and this is reasonable as expected. It is clear that the efficiency is close to 1 in most of the $\cos\theta$ region and $J/\psi$ mesons with smaller $|\mathrm{y}|$ have bigger probability of being valid event. There are actually more valid events than that by directly using the experimental detecting angular cut to $J/\psi$.

\section{Summary\label{sec:summary}}

In this paper, we have calculated the prompt $J/\psi$ photoproduction in $e^+e^-$ collider based on NRQCD at leading order, including the contributions from both direct and resolved photons. The color-octet channels dominate the total cross section. Only 5$\%$ of the $J/\psi$ mesons are from the decays of heavier charmonium. We present different kinematic distributions of both the cross section and the number of $J/\psi$ events. Under simple assumptions, we also investigated the issue of detecting efficiency. It shows that sizable $J/\psi$ photoproduction events could be collected. There are more than 50 events for $p_t=20\gev$ at the CEPC in comparison with only 16 $J/\psi$ events for $p_t>1\gev$ at LEP-II. It renders the $J/\psi$ photoprodution at the CEPC a well laboratory to test the color-octet mechanism in NRQCD. We suggest that $J/\psi$ photoproduction should be measured at future CEPC. The measurement will help to clarify the universality problem in NRQCD and improve our understanding of the quarkonium production mechanism.

\begin{acknowledgments}
	
This work was supported by the National Natural Science Foundation of China with Grant No. 11475183 and the Key Research Program of Frontier Sciences, CAS, Grant No. Y7292610K1.

\end{acknowledgments}



\section{Appendix}
As mentioned above we make two approximations or assumptions, i.e., $J/\psi$ mesons decays into $\mu^+\mu^-$ pair immediately after their production at the colliding point, and at the rest frame of $J/\psi$ the momentum distribution of $\mu^\pm$ is isotropic. 
Firstly at the rest frame of $J/\psi$, the 4-momenta of $\mu^+\mu^-$ pair in spherical coordinates are given as,
\begin{eqnarray}
\label{pmu}
p_{\mu^+}^*&=&(\frac{m_{J/\psi}}{2},\vec{p}^*_{\mu^+})\nonumber\\
&=&(\frac{m_{J/\psi}}{2},|\vec{p}^*_{\mu^+}|\sin\theta^* \cos\phi^*,|\vec{p}^*_{\mu^+}|\sin\theta^* \sin\phi^*,\nonumber\\
&&|\vec{p}^*_{\mu^+}|\cos\theta^*),\\
p_{\mu^-}^*&=&(\frac{m_{J/\psi}}{2},-\vec{p}^*_{\mu^+}),
\end{eqnarray}
where $\theta^*$ and $\phi^*$ are respectively the elevation angle and azimuth angle of $\vec{p}^*_{\mu^+}$, and $|\vec{p}_{\mu^+}^*|=\sqrt{\frac{1}{4}m_{J/\psi}^2 - m_{\mu^+}^2}$.

At laboratory frame, their parent $J/\psi$ meson has momentum $p_{J/\psi}$. We can make the Lorentz transformation of the $\mu^+\mu^-$ pair from rest frame of their parent $J/\psi$ to the laboratory frame. At laboratory frame, $\mu^{+(-)}$ has momentum $p_{\mu^{+(-)}}$ and taking the $e^+e^-$ beam direction as $z$-axis, the cosine of the angle between $\vec{p}_{\mu^{+(-)}}$ and $z$-axis are given as,
\begin{eqnarray}
\cos\theta_{\mu^{+(-)}}=\frac{\vec{p}_{\mu^{+(-)}z}}{\sqrt{E^2_{\mu^{+(-)}}-m^2_{\mu^{+(-)}}}}.
\end{eqnarray}

Now we can define the probability P of a $J/\psi$ meson to be detected,
\begin{eqnarray}
P=\frac{\int^1_{-1}\int^{2\pi}_0f(\cos\theta^*,\phi^*)d\cos\theta^*d\phi^*}{4\pi},
\end{eqnarray}
where $f(\cos\theta^*,\phi^*)=1$ for both $|\cos\theta_{\mu^+}|$ and $|\cos\theta_{\mu^-}| \geq 0.98$ and $f=0$ for others. Using numerical integration, the value of the probability of a $J/\psi$ meson with given momentum can be obtained.

\bibliography{jpsi}

\end{document}